\documentclass[aps,prl,superscriptaddress,showpacs,preprint,onecolumn]{revtex4}
 \usepackage{amsmath}  
 \usepackage{makeidx}
 \usepackage{graphicx}
 \usepackage{amsfonts}
 \usepackage[ansinew]{inputenc}
 \usepackage[usenames,dvipsnames]{pstricks}
 \usepackage{subfigure}
 \usepackage{mathrsfs}
 \allowdisplaybreaks[1]

              \makeindex

\begin{document}
\title{Alpha Decay to Doubly Magic Core in Quartetting Wave Function Approach}
\author{Shuo Yang}
\affiliation{School of Physics, Nanjing University, Nanjing 210093, China}
\author{Chang Xu}
\affiliation{School of Physics, Nanjing University, Nanjing 210093, China}
\author{Gerd R\"{o}pke}
\affiliation{Institut f\"{u}r Physik, Universit\"{a}t Rostock, D-18051 Rostock, Germany}
\affiliation{National Research Nuclear University (MEPhI), 115409 Moscow, Russia}
\author{Peter Schuck}
\affiliation{Institut de Physique Nucl\'{e}aire, Universit\'e Paris-Sud, IN2P3-CNRS, UMR 8608, F-91406, Orsay, France}
\affiliation{Univ. Grenoble Alpes, CNRS, LPMMC, 38000 Grenoble, France}
\author{Zhongzhou Ren}
\affiliation{School of Physics Science and Engineering, Tongji University, Shanghai 200092, China
}
\author{Yasuro Funaki}
\affiliation{Laboratory of Physics, Kanto Gakuin University, Yokohama 236-8501, Japan}
\author{Hisashi Horiuchi}
\affiliation {Research Center for Nuclear Physics (RCNP), Osaka University, Osaka  567-0047, Japan}
\affiliation {International Institute for Advanced Studies, Kizugawa 619-0225,  Japan}
\author{Akihiro Tohsaki}
\affiliation{Research Center for Nuclear Physics (RCNP), Osaka University, Osaka 567-0047, Japan}
\author{Taiichi Yamada}
\affiliation{Laboratory of Physics, Kanto Gakuin University, Yokohama 236-8501, Japan}
\author{Bo Zhou}
\affiliation{Faculty of Science, Hokkaido University, Sapporo 060-0810, Japan}

\date{\today}

\begin{abstract}
We present a microscopic calculation of $\alpha$-cluster formation in heavy nuclei $^{104}$Te ($\alpha$+$^{100}$Sn), $^{212}$Po ($\alpha$+$^{208}$Pb) and their neighbors $^{102}$Sn, $^{102}$Te, $^{210}$Pb and $^{210}$Po by using the quartetting wave function approach. Improving the local density approximation, the shell structure of the core nucleus is considered, and the center-of-mass (c.o.m.) effective potential for the quartet is obtained self-consistently from the shell model wavefunctions. The $\alpha$-cluster formation and decay probabilities are obtained by solving the bound-state of the c.o.m. motion of the quartet and the scattering state of the formed $\alpha$-cluster in the Gurvitz approach. Striking shell effects on the $\alpha$-cluster formation probabilities are analyzed for magic numbers 50, 82 and 126. The computed $\alpha$-decay half-lives of these special nuclei are compared with the newest experimental data.
\end{abstract}


\maketitle

\section{Introduction}

The $\alpha$-cluster formation problem is an important and challenging issue in not only light nuclei but also heavy and superheavy nuclei. Although much effort has been paid to it, this problem has still not been fully solved up to now \cite{Mang60,Delion09,DLW,Buck,Denisov,Mohr,cxu06,Royer}. For light nuclei, microscopic approaches have been successfully used to investigate the $\alpha$-like correlations (quartetting) with a full account of the Pauli exclusion principle \cite{THSR,RGM,GCM,FMD,AMD}. On the contrary, it is rather difficult to describe microscopically the formation of $\alpha$-clusters in heavy (superheavy) nuclei because it involves a complex many-body problem \cite{Po}. Reasonable approximations to the \emph{ab} initio methods should be adopted to make the calculations feasible within present computer capabilities \cite{Po}. An ideal heavy $\alpha$-emitter for testing those approximations is the nucleus $^{212}$Po, which is the instance of $\alpha$-decay to a doubly-magic core ($^{208}$Pb). Recently, another heavy $\alpha$-emitter $^{104}$Te was reported for the first time in experiment, which is not only the second instance of $\alpha$-decay to a doubly magic core ($^{100}$Sn) but also a self-conjugate nucleus \cite{Te2018}. Subsequent experimental search for $^{104}$Te observed two new events with properties consistent with the previously reported data \cite{Te2019}.

With the growth of new data in the past several years \cite{AME2016}, one great challenge for the field is to describe quantitatively $\alpha$-like correlations and its decay in heavy (superheavy) nuclei from first principles. So far, only a few microscopic works have been carried out to treat the $\alpha$-cluster formation problem in $^{212}$Po (see reviews \cite{review1,review2,review3}). Motivated by the concept of pairing in nuclei and inspired by the THSR (Tohsaki-Horiuchi-Schuck-R\"{o}pke) wave function concept that has been successfully applied to light nuclei \cite{THSR}, we recently proposed a Quartetting Wave Function Approach (QWFA) for describing $\alpha$-clustering and decay in heavy and superheavy nuclei \cite{Xu2016,Xu2017,quartet2017,quartet2018}. Unlike the light nuclei, the quartet consisting of n$\uparrow$, n$\downarrow$, p$\uparrow$, p$\downarrow$ ($\alpha$-like cluster) is considered to move with respect to a fixed center because the core nucleus is heavy, \textit{i.e.}, recoil effects are neglected. Even with this approximation where the core is considered as a mean field, the problem of $\alpha$-cluster formation is still very difficult to handle as one needs to treat correctly both the intrinsic motion between four nucleons in the cluster and the relative motion of the cluster versus the core \cite{Po}. The respective center of mass (c.o.m.) and intrinsic Schr\"{o}dinger equations are coupled in a complex way by contributions containing derivative terms of the intrinsic wave function with respect to the c.o.m. coordinate. No investigations of such derivative terms have performed yet for finite nuclear systems \cite{Po}. The solution of QWFA should join two limiting cases, the situation where the quartet is well inside the core nucleus and a shell-model calculation can be performed, and the limit of distant clusters. In particular, we consider that an $\alpha$-like state in QWFA can exist only at densities lower than the critical density $\rho_c$=0.2917 fm$^{-3}$ (see Refs.\cite{Po,quartet2017,quartet2018}) and dissolves at higher densities ($\rho>\rho_c$) into nearly uncorrelated (free) single-quasiparticle states due to self-energy shifts and the Pauli blocking effects. By using the two-potential technique \cite{Gurvitz1988}, the $\alpha$-cluster formation and decay probabilities can be well defined by solving the bound-state of the c.o.m. motion of the quartet and the scattering state of the formed $\alpha$-cluster \cite{Xu2016,Xu2017}.

Within the local density approximation (LDA) (strictly valid for infinite matter), the Thomas-Fermi model for the core nucleus was taken in our previous calculations \cite{Xu2016,Xu2017}. According to the Thomas-Fermi rule \cite{quartet2017,quartet2018}, four nucleons are added to the core at the sum of the respective Fermi energies which is identical with the tunneling energy of the emitted $\alpha$-particle (bound state energy -28.3 MeV plus the kinetic energy which gives together the tunneling energy, in the case of $^{212}$Po: -19.52 MeV). However, this approach is not fully consistent. In particular, it is not able to describe the nuclear shell structure of the core nucleus. This rule is too restrictive, and nucleons are added to the core in shell states which have a finite energy difference above occupied states in the core. A gap in the strength of nucleon-nucleon (NN) interaction has been found through the fitting of realistic $\alpha$-decay lifetimes of Po isotopes \cite{Batchelder}. In this work, we investigate the c.o.m. motion of $\alpha$-like quartet moving under the shell structure influence of core nucleus. We improve the Thomas-Fermi rule by introducing quasiparticle (shell model) nucleon states for the core nucleus. In contrast to former investigations in Refs.\cite{Xu2016,Xu2017}, the c.o.m. effective potential for the quartet is now obtained self-consistently from the contributing shell model wavefunctions with the same NN interaction.

We perform calculations of both $\alpha$-cluster formation and decay probabilities in ideal heavy $\alpha$-emitters $^{104}$Te, $^{212}$Po and their neighbors $^{102}$Sn, $^{102}$Te, $^{210}$Pb and $^{210}$Po. Comparisons of the c.o.m. effective potentials and quartetting wave functions are made between neighboring nuclei. The underlying physics of striking structure effects across major shells 50, 82, and 126 on the $\alpha$-cluster formation and decay probabilities are analyzed in detail.

This paper is organized as follows. In Section 2, the formulism of coupled intrinsic and c.o.m. Schr\"{o}dinger equations of the quartet is explicitly given. The shell model wave functions of the quartet nucleons are displayed in Section 3. The c.o.m. effective potential of quartet is discussed in Section 4. Section 5 gives the numerical results of $\alpha$-cluster formation and decay probabilities from QWFA. The last section gives a short summary.

\section{Intrinsic and C.O.M. Schr\"{o}dinger Equations of Quartet}

The main ingredient of the quartetting wave function approach is the introduction of a collective variable $\bf R$, describing the c.o.m. motion of the quartet, and variables that describe the intrinsic motion ${\bf s}_j=\{\bf S,s,s'\}$ with the Jacobi-Moshinsky coordinates for the quartet nucleons \cite{Po,quartet2017,quartet2018}:
\begin{eqnarray}
\label{Jacobi}
&&{\bf r}_{n,\uparrow}={\bf R}+{\bf S}/2+{\bf s}/2,\,\,\,\,{\bf r}_{n,\downarrow}={\bf R}+{\bf S}/2-{\bf s}/2,\nonumber\\
&&{\bf r}_{p,\uparrow}={\bf R}-{\bf S}/2+{\bf s'}/2,\,\,\,{\bf r}_{p,\downarrow}={\bf R}-{\bf S}/2-{\bf s'}/2.
\end{eqnarray}

The energy eigenstate $\Phi ({\bf R},{\bf s_j})$ of the quartet can be subdivided in a unique way into a c.o.m. motion part $\Psi ({\bf R})$ and an intrinsic motion part $\varphi ^{\rm intr}({\bf s_j},{\bf R})$
\begin{eqnarray}
\label{subd}
\Phi ({\bf R},{\bf s_j})=\varphi ^{\rm intr}({\bf s_j},{\bf R})\Psi ({\bf R})
\end{eqnarray}
with the normalization condition
\begin{eqnarray}
\label{normalization}
\int d^{3}R\int d^{9}s_j \left | \Phi ({\bf R},{\bf s_j}) \right |^2=1,\,\,\,\,\int d^{9}s_j \left | \varphi ^{\rm intr} ({\bf s_j},{\bf R}) \right |^2=1.
\end{eqnarray}

The Hamiltonian of the $\alpha$-cluster can be written as
\begin{equation}
\label{Hc}
H=\left(-\frac{\hbar^2}{8m} \nabla_R^2+T[\nabla_{s_j}]\right)\delta^3({\bf R}-{\bf R}')\delta^3({\bf s}_j-{\bf s}'_j)
+V({\bf R},{\bf s}_j;{\bf R}',{\bf s}'_j)
\end{equation}
where $-\frac{\hbar^2}{8m} \nabla_R^2$ is the kinetic energy of the c.o.m. motion and $T[\nabla_{s_j}]$ the kinetic energy of the internal motion of the quartet. The interaction $V({\bf R},{\bf s}_j;{\bf R}',{\bf s}'_j)$ contains the mutual interaction between quartet nucleons as well as the interaction of the quartet nucleons with an external potential. For the c.o.m. motion of the quartet we have the Schr\"{o}dinger equation
\begin{eqnarray}
\label{comeq}
&&-\frac{\hbar^2}{8m} \nabla_R^2\psi({\bf R})-\frac{\hbar^2}{Am}\int d^9s_j \varphi^{{\rm intr},*}({\bf s}_j,{\bf R})
[\nabla_R \varphi^{{\rm intr}}({\bf s}_j,{\bf R})][\nabla_R\psi({\bf R})]
\nonumber\\ &&
-\frac{\hbar^2}{8m}\int\!\! d^9s_j \varphi^{{\rm intr},*}({\bf s}_j,{\bf R})
[ \nabla_R^2 \varphi^{{\rm intr}}({\bf s}_j,{\bf R})] \psi({\bf R})
+\!\!\int \!\! d^3R'\,W({\bf R},{\bf R}')  \psi({\bf R}')\!=\!E\,\psi({\bf R}),
\end{eqnarray}
with the c.o.m. potential
\begin{eqnarray}
\label{compo}
W({\bf R},{\bf R}')&=&\int d^9s_j\,d^9s'_j\,\varphi^{{\rm intr},*}({\bf s}_j,{\bf R}) \left[T[\nabla_{s_j}]
\delta^3({\bf R}-{\bf R}')\delta^9({\bf s}_j-{\bf s}'_j)\right.\nonumber \\&&\left.
+V({\bf R},{\bf s}_j;{\bf R}',{\bf s}'_j)\right]
\varphi^{{\rm intr}}({\bf s}'_j,{\bf R}').
\end{eqnarray}
For the intrinsic motion of quartet nucleons we have another Schr\"{o}dinger equation
\begin{eqnarray}
\label{intr}
&&-\frac{\hbar^2}{4m}  \psi^*({\bf R}) [\nabla_R\psi({\bf R})]
[\nabla_R \varphi^{{\rm intr}}({\bf s}_j,{\bf R})]
-\frac{\hbar^2}{8m}  |\psi({\bf R})|^2
\nabla_R^2 \varphi^{{\rm intr}}({\bf s}_j,{\bf R})
\nonumber \\ &&
+\int d^3R'\,d^9s'_j\, \psi^*({\bf R}) \left[T[\nabla_{s_j}]
\delta^3({\bf R}-{\bf R}')\delta^9({\bf s}_j-{\bf s}'_j)\right.\nonumber \\&& \left.
+V({\bf R},{\bf s}_j;{\bf R}',{\bf s}'_j)\right]
 \psi({\bf R}')\varphi^{{\rm intr}}({\bf s}'_j,{\bf R}')=F({\bf R}) \varphi^{{\rm intr}}({\bf s}_j,{\bf R}).
\end{eqnarray}
The respective c.o.m. and intrinsic Schr\"{o}dinger equations are coupled by contributions containing the expression $\nabla_R \varphi^{{\rm intr}}({\bf s}_j,{\bf R})$ which disappears in homogeneous matter. The approach presented here to include four-nucleon correlations is based on a first-principle approach to nuclear many-body  systems, however, several approximations should be performed to make the approach practicable. One of the approximations is that the derivative terms  $\nabla_R \varphi^{\rm intr}({\bf s_j},{\bf R})$ in Eqs.(\ref{comeq}) and (\ref{intr}) are not included in QWFA at present.

\section{Shell Model Wave Functions of Quartet Nucleons}

\begin{figure}[htb]
\includegraphics[width=0.85\textwidth]{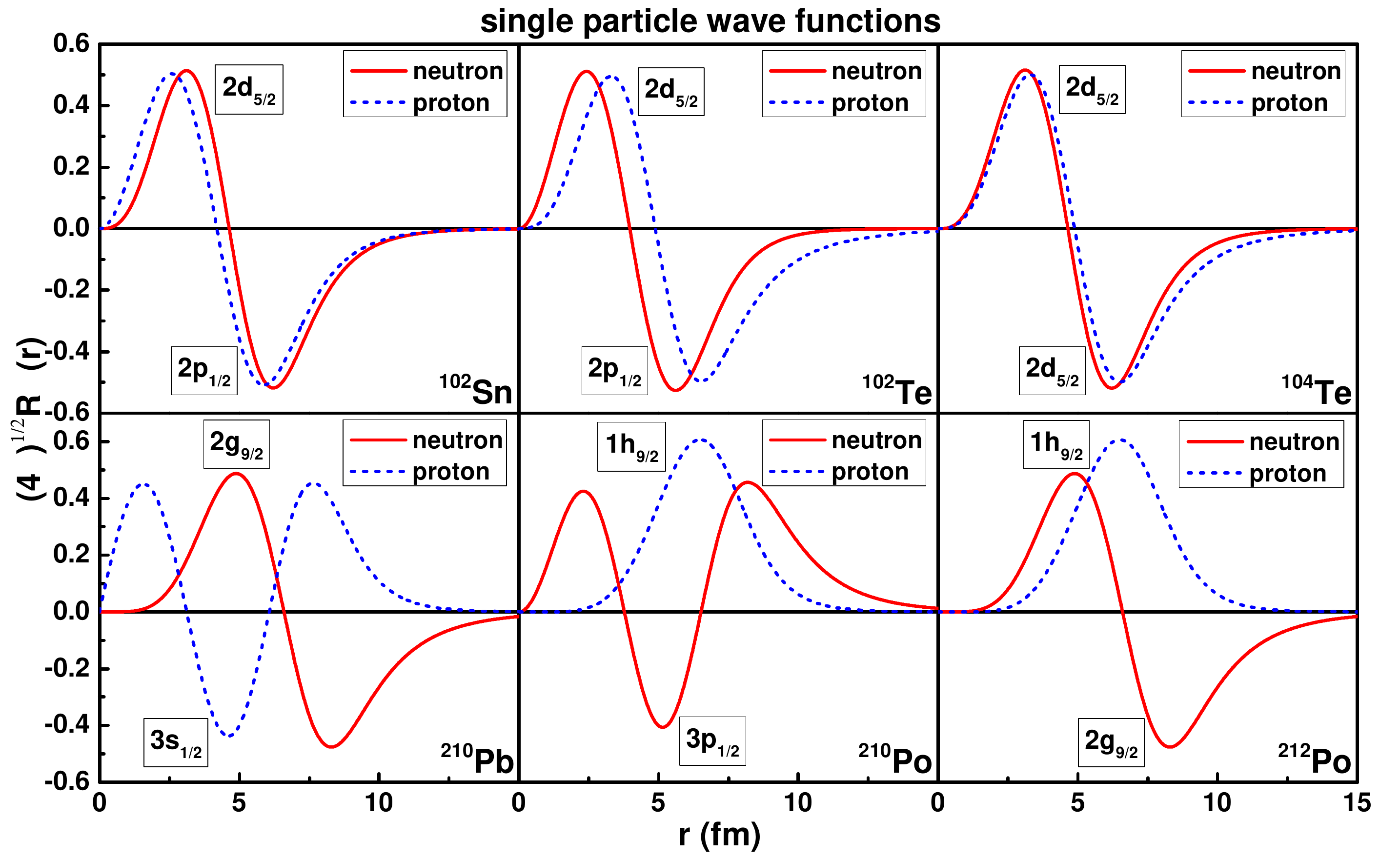}
\caption{(Color online) The contributing single-particle wave functions of protons and neutrons in the quartets of $^{102}$Sn, $^{102}$Te, $^{104}$Te, $^{210}$Pb, $^{210}$Po and $^{212}$Po.}
\label{Fig:1}
\end{figure}

In previous calculations, the core nucleons have been treated within the Thomas-Fermi approximation as the simplest version of LDA \cite{Xu2016,Xu2017}. To introduce a quartet at minimum energy, each nucleon must be added at the corresponding Fermi energy. From this, two consequences follow immediately: First, the effective potential inside the core for the c.o.m. motion of the quartet is constant given by the constant chemical potential, not strongly increasing as usually found in the literature, and second that the value of the sum of these four chemical potentials coincides with the energy of the emitted $\alpha$-particle. This Thomas-Fermi rule for the core nucleus was assumed in QWFA. This rule is quite simple and gives a local density description for the quartet, which is not able to describe self-consistently the shell effects as observed in our previous work (Ref.\cite{Xu2017}). Here we improve the Thomas-Fermi rule by taking the discrete level structure of the core nucleus into account. We use the Woods-Saxon potential + $ls$ coupling to determine single-nucleon orbits that are occupied up to the
Fermi energy \cite{SPM}
\begin{equation}
\label{ws1}
V_{WS}\left ( r \right ) = \frac{V_{0}}{1+\textrm{exp}(\frac{r-R_0}{a})},
\end{equation}
where the strength of the Woods-Saxon potential is parameterized as $V_{0}=-46\left [ 1\pm 0.97( \frac{N-Z}{A} ) \right ]$(``$+$" for protons and ``$-$" for neutrons). The parameter $R_0$ is $1.43A^{1/3}$ fm for both protons and neutrons and the diffusivity parameter $a$ is 0.7 fm. The Coulomb potential we adopt is
\begin{eqnarray}
\label{ws2}
V_C\left ( r \right )=(Z-1)e^2\begin{cases} (3R_{\rm Coul}^2-r^2)/2R_{\rm Coul}^3,&r\le  R_{\rm Coul} \\1/r,&r> R_{\rm Coul}\end{cases}
\end{eqnarray}
with the radius $R_{\rm Coul}=1.25A^{1/3}$ fm. For the $ls$ coupling potential, we use the following form
\begin{equation}
\label{ws3}
V_{\rm so}\left ( r \right ) = \frac{1}{2\mu^2 r}\left ( \frac{\partial }{\partial r} \frac{\lambda V_{0}}{1+\textrm{exp}(\frac{r-R_{\rm so}}{a_{\rm so}})}\right ) \bf l \cdot \bf s,
\end{equation}
where $\mu$ is the reduced mass of the $\alpha$-core system. The diffusivity parameter $a_{\rm so}$ is also 0.7 fm and the parameter $R_{\rm so}$ is $1.37A^{1/3}$ fm. The normalization factor of the $ls$ coupling strength $\lambda$ is 37.5 for neutrons and 31 for protons, respectively. Note that other choices of parameter sets for the Woods-Saxon potential + $ls$ coupling can be used in QWFA, provided that the shell model states are correctly reproduced. In Fig.\ref{Fig:1}, the contributing single-particle wave functions of the quartet are shown for the $\alpha$-emitters $^{104}$Te, $^{212}$Po and their neighbors $^{102}$Sn, $^{102}$Te, $^{210}$Pb and $^{210}$Po. It is clear that only states near the Fermi energy can form an $\alpha$-like cluster because only these single-particle states extend to the low-density regions at the surface of the nucleus \cite{Xu2016,Xu2017}. The quartet will be introduced on top of the core nucleus in the shell above the Fermi level. As the $^{104}$Te is a self-conjugate nucleus, the protons and neutrons in the quartet occupy the same $2d_{5/2}$ single-particle states. This is contrary to the case of $^{212}$Po where the proton and neutron orbits in the quartet are quite different (see Fig.\ref{Fig:1}). A problem to be solved in future investigations is the treatment of partially filled shells, when spherical symmetry can not be assumed.

\section{C.O.M. Effective Potential of Quartet}

The main issue in this section is to obtain an effective potential for the c.o.m. motion of the quartet from the contributing single-particle wave functions shown in previous section. In particular, the behavior of the effective potential inside the core is of interest. The Thomas-Fermi model demands a constant behavior, and previous calculations with nucleon orbitals \cite{quartet2017,quartet2018} show also a nearly constant behavior. Here we derive results using realistic shell-model states. The quartet wave function $\Phi _{4}$ with the Jacobi-Moshinsky coordinates $({\bf R,S,s,s'})$ is given by:
\begin{eqnarray}
\label{scquartet1}
\Phi _{4}({\bf R,S,s,s'})=&&\Phi({\bf r_{1}},{\bf \sigma_{1}};{\bf r_{2}},{\bf \sigma_{2}};{\bf r_{3}},{\bf \sigma_{3}};{\bf r_{4}},{\bf \sigma_{4}})\nonumber\\
=&&\sum_{J_{12},M_{12},J_{34},M_{34}}\left \langle J_{12},M_{12},J_{34},M_{34}|J,M \right \rangle\sum _{m_{1},m_{2}}\left \langle j_{1},m_{1},j_{2},m_{2}|J_{12},M_{12} \right \rangle\nonumber\\
&&\left |j_{1},m_{1} \right \rangle \left |j_{2},m_{2} \right \rangle\sum _{m_{3},m_{4}}\left \langle j_{3},m_{3},j_{4},m_{4}|J_{34},M_{34} \right \rangle \left |j_{3},m_{3} \right \rangle\left |j_{4},m_{4}\right \rangle,
\end{eqnarray}
where the notations $1$ and $2$ denote two protons in the quartet, and $3$ and $4$ two neutrons. The quantum numbers for the total angular momentum and its z-component of nucleon $i$ are denoted by $j_{i}$ and $m_{i}$, respectively. $j_1$ and $j_2$ are coupled to $J_{12}$, $j_3$ and $j_4$ to $J_{34}$, and then $J_{12}$ and $J_{34}$ to $J$. Here we consider only the ground state $\alpha$-transitions of even-even nuclei, so that we have $J_{12}=J_{34}=J=0, M_{12}=M_{34}=M=0$. Therefore, the quartet wave function can be subdivided into the wave function of two protons $\Phi({\bf r_{1}},{\bf \sigma_{1}};{\bf r_{2}},{\bf \sigma_{2}})$ and the wave function of two neutrons $\Phi({\bf r_{3}},{\bf \sigma_{3}};{\bf r_{4}},{\bf \sigma_{4}})$. Let $a,b=1,2$ or $3,4$, the Fourier transformation of the wave function of two nucleons $\Phi({\bf r_{a}},{\bf \sigma_{a}};{\bf r_{b}},{\bf \sigma_{b}})$ is
\begin{eqnarray}
\label{sc phi_p2}
\varphi _{ab}({\bf p})=&&\frac{1}{(2\pi )^6}\int d^3{r_a }\int d^3{r_b }\left | \Phi({\bf r_{a}},{\bf \sigma_{a}};{\bf r_{b}},{\bf \sigma_{b}}) \right |^2e^{-i{\bf p\cdot r_a}-i{\bf p\cdot r_b}}\nonumber\\
=&&\frac{1}{(2\pi )^6}\sum _{m_{sa},m_{sb}}\sum _{m_{a},m'_{a},m_{b},m'_{b}}\sum _{m_{la},m'_{la},m_{lb},m'_{lb}}\left \langle j_{a},m_{a},j_{b},m_{b}|0,0 \right \rangle \left \langle j_{a},m'_{a},j_{b},m'_{b}|0,0 \right \rangle\nonumber\\
&&\left \langle l_{a},m_{la},\frac{1}{2},m_{sa}|j_{a},m_{a} \right \rangle\left \langle l_{b},m_{lb},\frac{1}{2},m_{sb}|j_{b},m_{b} \right \rangle\left \langle l_{a},m'_{la},\frac{1}{2},m_{sa}|j_{a},m'_{a} \right \rangle\nonumber\\
&&\left \langle l_{b},m'_{lb},\frac{1}{2},m_{sb}|j_{b},m'_{b} \right \rangle f_{l_a,m_{la},m'_{la}}\left (\bf p\right ) f_{l_b,m_{lb},m'_{lb}}\left (\bf p\right ),
\end{eqnarray}
where the function $f_{l,m,m'}(\bf p)$ can be obtained from the contributing single-nucleon wave functions
\begin{eqnarray}
\label{intergral}
f_{l,m,m'}(\bf p)=&&\int d^3{r}R_{nl}^2(r)Y_{lm}^*(\theta_r,\varphi_r)Y_{lm'}(\theta_r,\varphi_r)e^{-i{\bf p}\cdot {\bf r}}\nonumber\\
=&&4\pi \sum_{l'=0}^{2l}(-i)^{l'}\sqrt{\frac{2l'+1}{4\pi }}\left \langle l,0,l',0|l,0 \right \rangle \left \langle l,m,l',m'-m|l,m'\right \rangle Y_{l'm'-m}(\theta_p,\varphi_p)\nonumber\\
&&\int_{0}^{\infty } r^2R_{nl}^2(r)j_{l'}(pr)dr.
\end{eqnarray}
The density distribution of the quartet $\rho _{4}(R)$ is then given by
\begin{eqnarray}
\label{phi_p2}
\rho _{4}(R)=&& \int d^3{S}d^3{s}d^3{s'}|\Phi_{4}({\bf R, S, s, s}')|^2\\ \nonumber
            =&& 2^6(2\pi )^9\int d^3{p}\varphi _{12}({\bf p})\varphi _{34}({\bf p})e^{4i{\bf p}\cdot{\bf R}}.
\end{eqnarray}
The wave function $\Psi_{4}({R})=\rho _{4}^{1/2}(R)$ of the c.o.m. motion of the quartet which corresponds to the density distribution $\rho _{4}(R)$ follows the Schr\"{o}dinger equation \cite{quartet2017}:
\begin{equation}
\label{W1}
-\frac{\hbar^2}{8 m} \nabla_R^2 \Psi_{4}( R)+W( R) \Psi_{4}( R)=E  \Psi_{4}( R).
\end{equation}
Let us introduce $u_{4}(R)=(4\pi)^{1/2}R\Psi_{4}(R)$, the effective c.o.m potential $W{(R)}$ of the quartet is then given by:
\begin{equation}
\label{W2}
W(R)-E=\frac{\hbar^2}{8m} \frac{u_{4}''( R)}{u_{4}( R)}=\frac{\hbar^2}{8m}\frac{\rho_{4} '(R)}{R\rho_{4} (R)}-\frac{\hbar^2}{32m}\frac{\rho_{4} '(R)^2}{\rho_{4} (R)^2}+\frac{\hbar^2}{16m}\frac{\rho_{4} ''(R)}{\rho_{4} (R)}.
\end{equation}

\begin{figure}[h]
\subfigure{\includegraphics[width=0.6\textwidth]{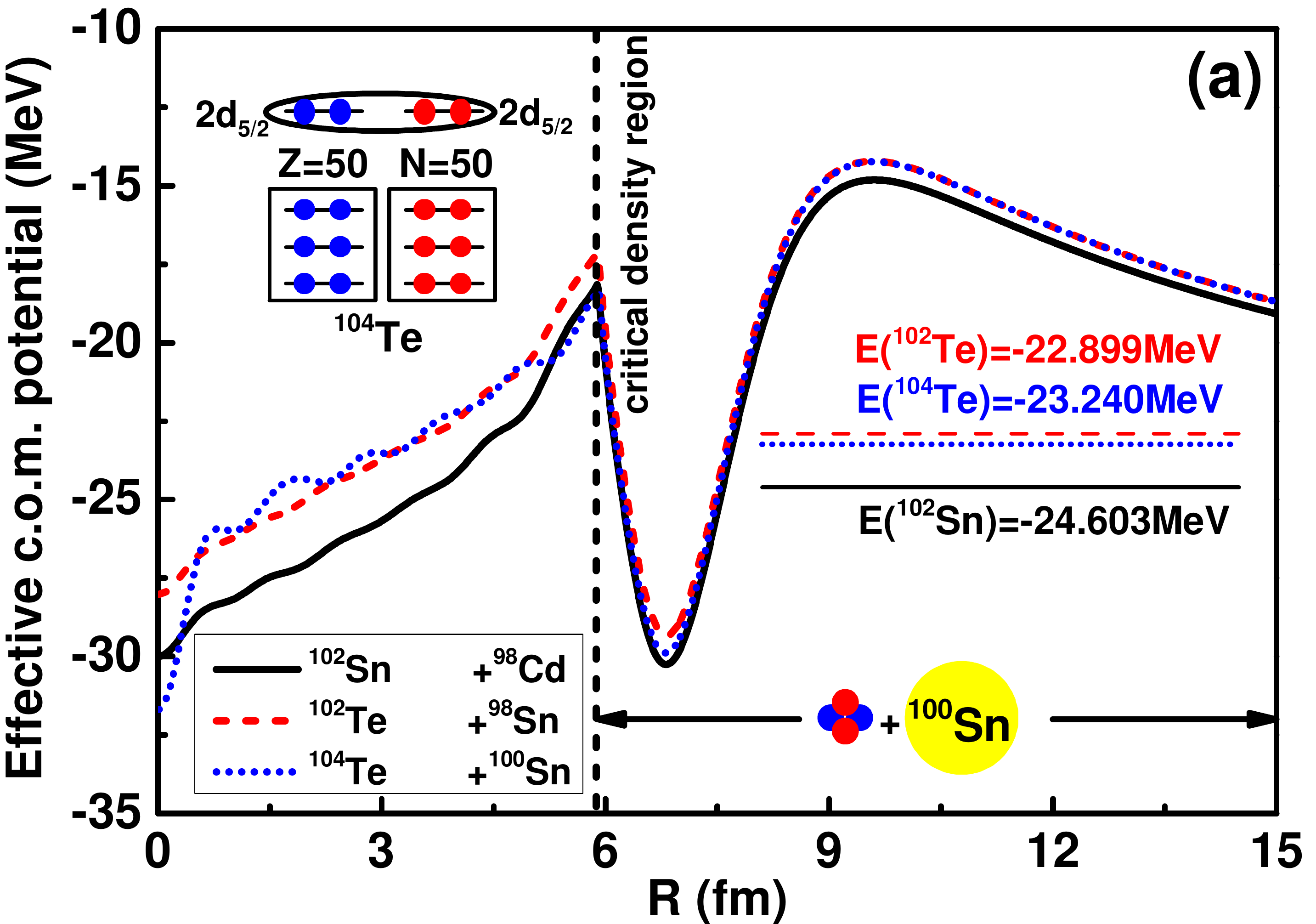}}
\subfigure{\includegraphics[width=0.6\textwidth]{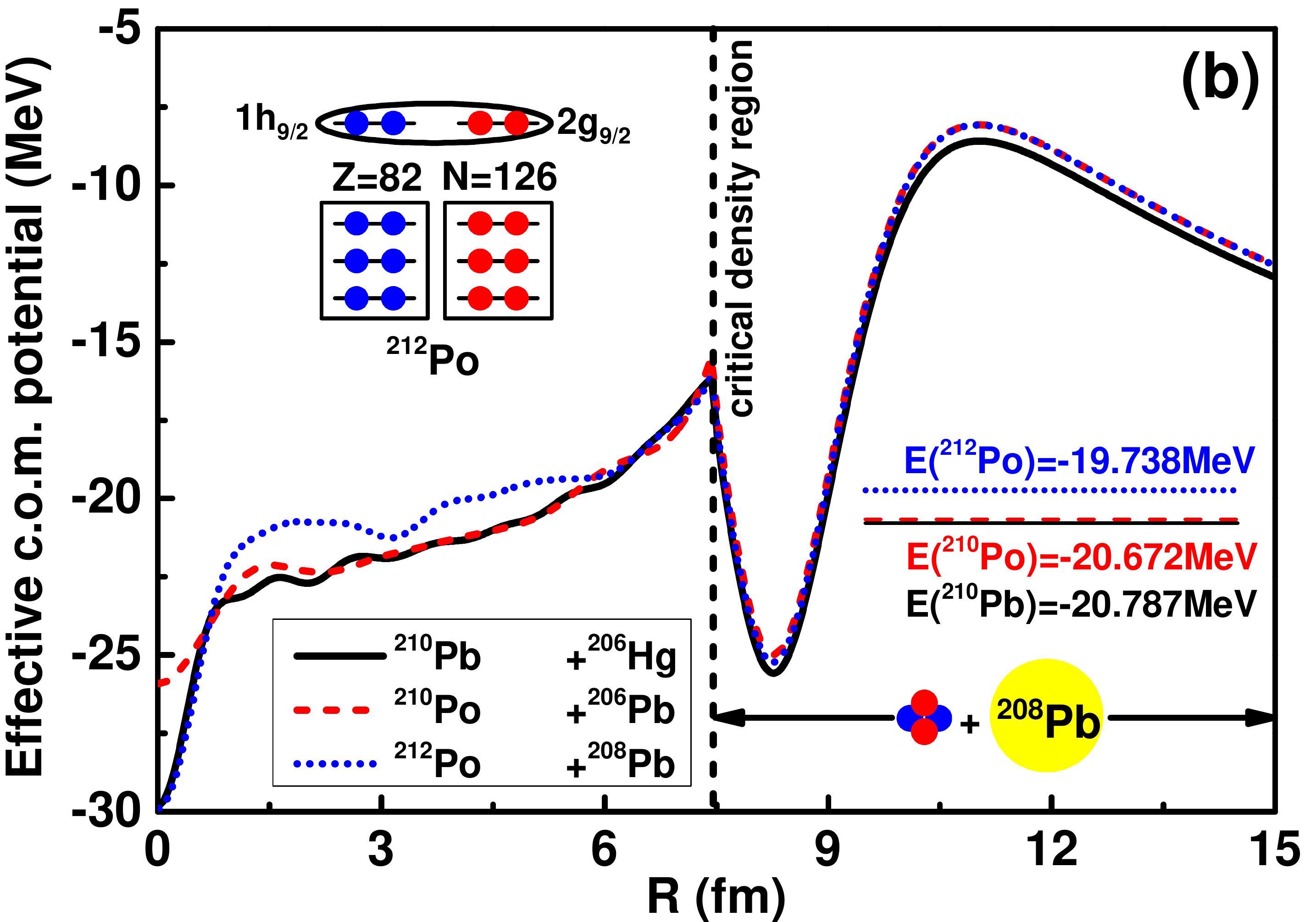}}
\caption{(Color online) The c.o.m. effective potentials of $\alpha$-cluster in (a) $^{102}$Sn, $^{102}$Te and $^{104}$Te, (b) $^{210}$Pb, $^{210}$Po and $^{212}$Po. The sketch of small box with filled circles denotes the core nucleus considered as a mean-field. The position of the energy eigenvalues $E$ for the c.o.m. motion is marked.}
\label{Fig:2}
\end{figure}

Note that our local potential $W(R)$ has the correct asymptotic behavior to the Coulomb potential at large distances, but inside the core where the Pauli principle acts, this local effective potential is expected to have wiggles (in contrast to the Thomas-Fermi model where it is constant). For the harmonic oscillator basis, the energy eigenvalue $E$ of the c.o.m motion of the quartet can be obtained easily by subtracting the intrinsic motion energy of the quartet from the sum of four single-particle energies \cite{quartet2017}. However, it is not easy to do so in the \textit{WS}+\textit{ls} coupled basis. Instead of subtracting the intrinsic motion energy from total energy,  here we join the effective c.o.m potential $W{(R)}$ in the core region smoothly with the $\alpha$-core interaction in surface region at the critical density $\rho_c$ \cite{Xu2016,Xu2017}. We use the following neutron and proton densities for the core nucleus:
\begin{eqnarray}
\label{Nnp}
\rho_n( R)=\frac{\rho_{n0}}{[1+e^{(R-R_{n0})/a_n}]}, \,\,\,\rho_p( R)=\frac{\rho_{p0}}{[1+e^{(R-R_{p0})/a_p}]}.
\end{eqnarray}
where the detailed values of the half-density radius and diffuseness parameter are given in Table~\ref{Tab:1}. After getting the baryon density $\rho_B=\rho_n+\rho_p$, we determine a critical radius $R_c$ corresponding to the critical density $\rho_c=0.02917$ fm$^{-3}$ for each nucleus. The critical radii are $R_c(^{98}$Cd$)=5.899$ fm, $R_c(^{98}$Sn$)=5.900$ fm, $R_c(^{100}$Sn$)=5.912$ fm, $R_c(^{206}$Hg$)=7.433$ fm, $R_c(^{206}$Pb$)=7.432$ fm and $R_c(^{208}$Pb$)=7.438$ fm, respectively. In the local density approximation, the formation/dissolution of the $\alpha$-cluster happens sharply at the critical radius $R_c$. At distances larger than $R_c$, there is a certain probability that an $\alpha$-cluster can be formed. The $\alpha$-core interaction $V(R)$ in surface region with $R>R_c$ consists of the attractive nuclear potential $V_N(R)$, the Coulomb potential $V_C(R)$ and the repulsive Pauli potential as a consequence of antisymmetrization between the $\alpha$-cluster and the core. The Pauli blocking term depends on the baryon density $\rho_B$ \cite{Po,quartet2017,quartet2018} as we use for the local density approximation
\begin{equation}
\label{WPauli}
 W^{\rm Pauli}(\rho_B)\approx 4515.9\, {\rm MeV\, fm}^3 \rho_B -100935\, {\rm MeV\, fm}^6 \rho_B^2+1202538\, {\rm MeV\, fm}^9 \rho_B^3.
\end{equation}

For the nuclear potential, the M3Y-type nucleon-nucleon interaction is used in the double folding procedure with matter density distributions of both $\alpha$ and core nucleus. This M3Y-type NN interaction consists of a short-range repulsion part and a long-range attraction part \cite{M3YReview},
\begin{eqnarray}
\label{M3Y}
V_N(R)=c \exp(-4R)/(4R)-d \exp(-2.5R)/(2.5R).
\end{eqnarray}

\begin{table*}[htb]
\caption{Parameters of the density distributions in the core nucleus, which are chosen based on the results in Refs. \cite{Tarbert2014,Fricke1995,r100,Warda}.}\label{Tab:1}
\begin{tabular}{|c|c|c|c|c|}
\hline
Core nucleus & $R_{n0}$ (fm) & $a_n$ (fm) & $R_{p0}$ (fm) & $a_p$ (fm)\\
\hline
$^{98}$Cd, $^{98}$Sn, $^{100}$Sn & 5.15 & 0.49 & 5.15 & 0.53\\
\hline
$^{206}$Hg, $^{206}$Pb, $^{208}$Pb & 6.70 & 0.55 & 6.68 & 0.447\\
\hline
\end{tabular}
\end{table*}

Previously, the strength parameters $c$ and $d$ of the NN interaction in Eq.(\ref{M3Y}) were adjusted for each $\alpha$-emitter by fitting the measured decay energy and half-life \cite{Xu2016,Xu2017}. Shell structure effect of the core nucleus manifests itself in the strength parameters and a gap in $c$ and $d$ values was observed \cite{Xu2017}. This is not satisfactory as one should in principle start from the same NN interaction instead of adjusting $c$ and $d$ for each nucleus. By replacing the Thomas-Fermi rule with the shell model calculations for the core nucleus, we are able to use the same strength parameters $c=17692$ and $d=4980$ for all $\alpha$-emitters considered here. Fig.\ref{Fig:2} exhibits the complete c.o.m effective potentials for these $\alpha$-emitters by joining smoothly the inner part $W{(R)}$ and outer part $V(R)$. With the Thomas-Fermi rule, the quartet c.o.m. motion inside the core nucleus was described by a constant effective potential, which is absolutely valid only for nuclear matter. As expected, it is observed here from Fig.\ref{Fig:2} that the effective c.o.m. potential behaves much more flat inside the heavy core $^{208}$Pb as compared to $^{100}$Sn. It is also observed that the c.o.m. potential inside core nucleus is quite sensitive to the details of contributing shell model states. A pocket is still formed for the effective potentials after introducing shell model states for the core nucleus. The sharp edge near the pocket is possibly a consequence of our approximation where the Pauli blocking shift of the $\alpha$-like cluster is given by the local density. Within a more detailed calculation taking into account the extension of the $\alpha$-like cluster and the nonlocal character of the Pauli blocking, we expect that this sharp edge will be washed out.

\section{Alpha Cluster Formation and Decay in Heavy Nuclei}

\begin{figure}[htb]
\subfigure{
\includegraphics[width=0.6\textwidth]{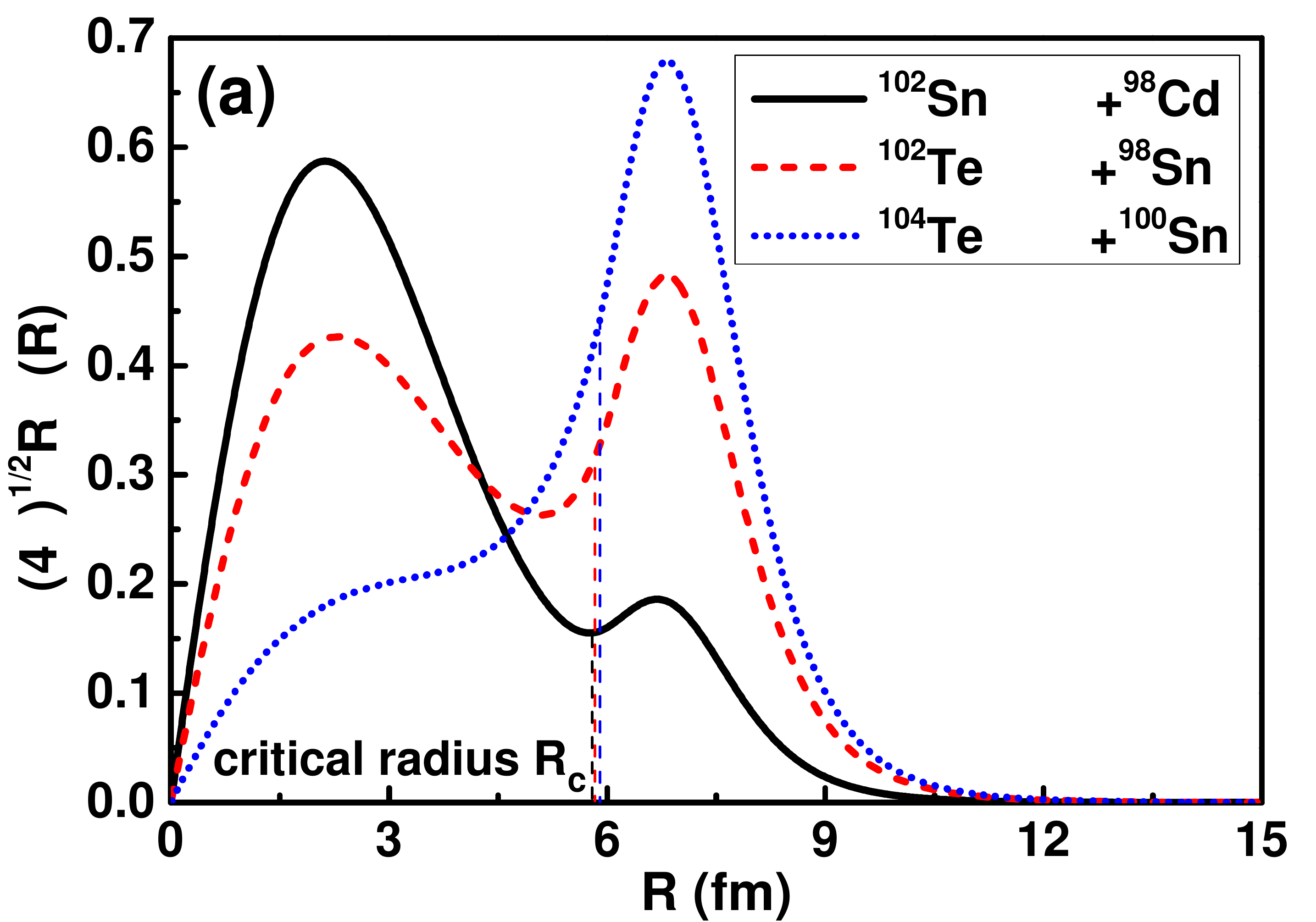}}
\subfigure{
\includegraphics[width=0.6\textwidth]{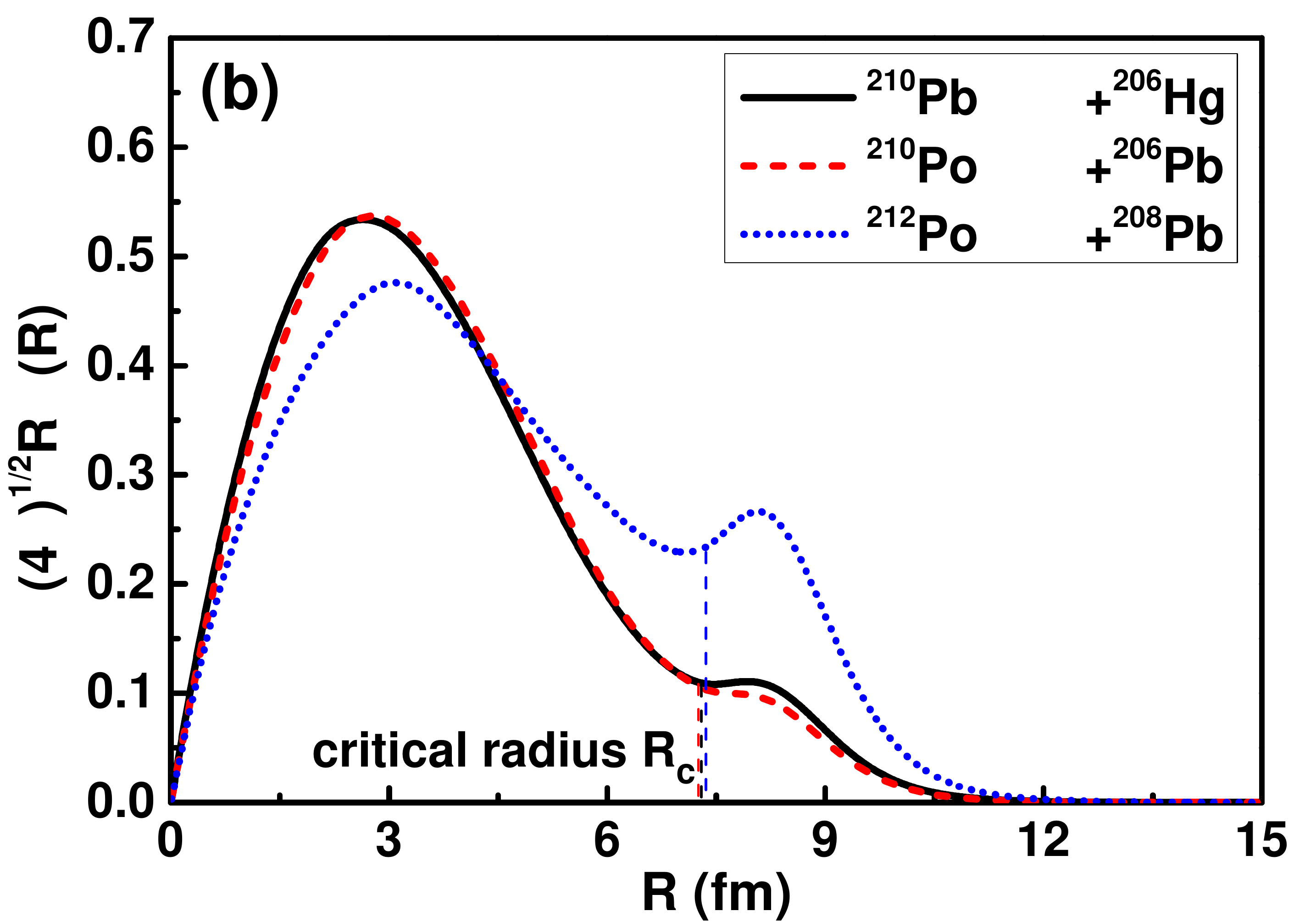}}
\caption{(Color online) Comparison of the normalized bound state wave functions for (a) $^{102}$Sn, $^{102}$Te and $^{104}$Te, (b) $^{210}$Pb, $^{210}$Po and $^{212}$Po. The positions of critical radii $R_c$ for each nucleus are marked. The two peaks or the shift of the maximum are caused by the formation of a pocket in Fig.\ref{Fig:2}.}
\label{Fig:3}
\end{figure}

\begin{figure}[htb]
\includegraphics[width=0.6\textwidth]{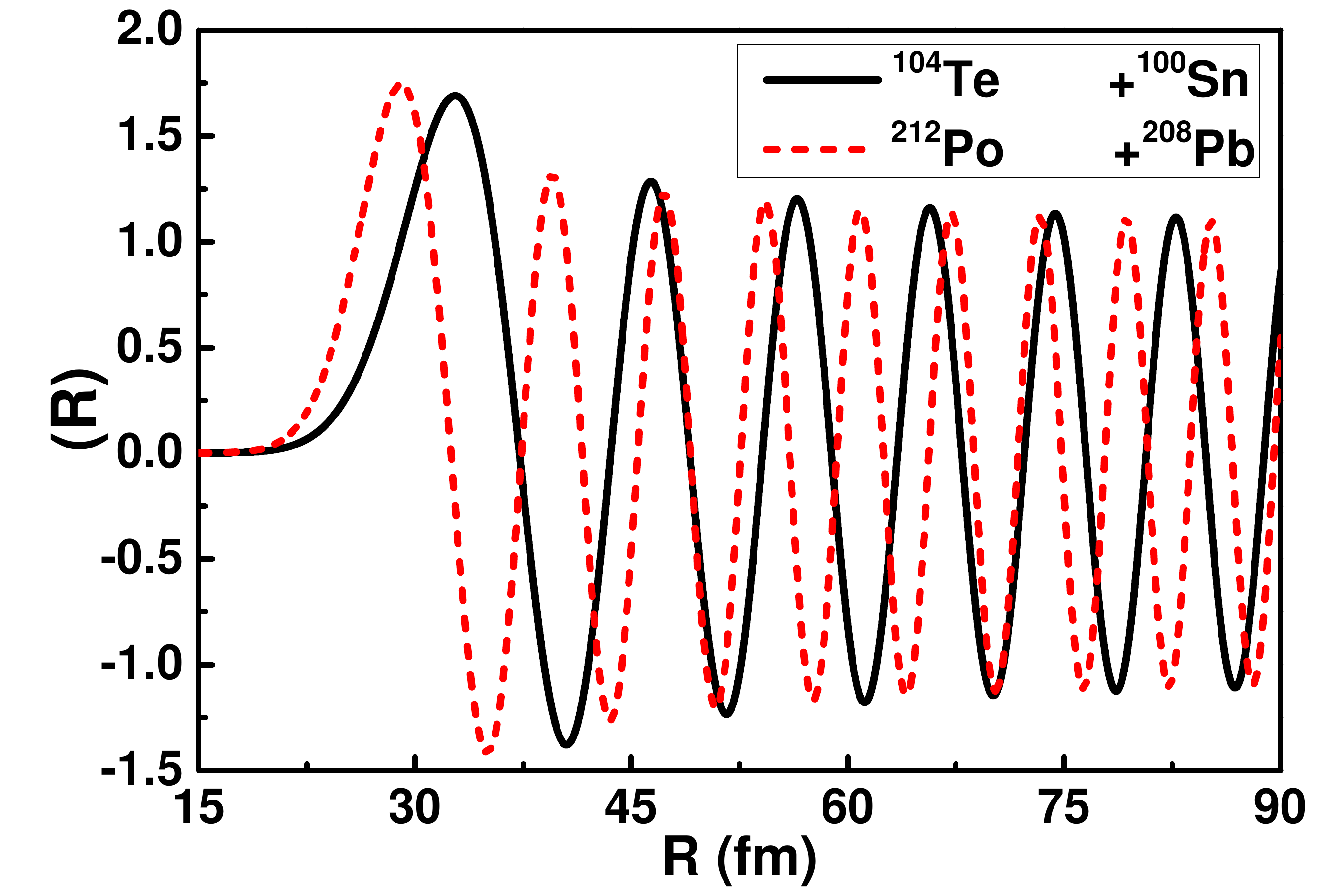}
\caption{(Color online) The scattering wave functions $\chi(R)$ for $\alpha$-emitters $^{104}$Te and $^{212}$Po in the two-potential approach. The separating point is chosen to be $R_{\rm sep} = 15$ fm.}
\label{Fig:4}
\end{figure}

Using the two-potential approach of Gurvitz \cite{Gurvitz1988}, the complete effective c.o.m. potential is split into two potentials at a separation radius $R_{\rm sep}$. The choice of separation radius does not affect the final result as long as it is large enough \emph{e.g.} $R_{\rm sep}$=15 fm. This method enables one to obtain a perturbative expansion for the decay width and the energy shift of a quasi-stationary state like the $\alpha$-decay. Both the bound state wave function $\Phi(R)$ of the first potential and the scattering state wave function $\chi(R)$ of the second one are calculated by solving the corresponding Schr\"{o}dinger equations. The normalized bound state c.o.m. wave functions $(4\pi)^{1/2}R\Phi(R)$ are plotted in Fig.\ref{Fig:3}. As clearly shown in Fig.\ref{Fig:3}, the bound-state wave functions $\Phi(R)$ of both $^{104}$Te and $^{212}$Po have a large component in the surface region with $R>R_c$ as compared with their neighbors. This is clearly due to the shell structure effect of core nucleus. The $\alpha$-cluster preformation probability $P_{\alpha}$ is obtained by integrating the bound-state wave function $\Phi(R)$ from the critical radius $R_c$ to infinity \cite{Po,Xu2016,Xu2017}:
\begin{eqnarray}
\label{palpha}
P_{\alpha}=\int_0^\infty  d^3R |\Phi(R)|^2 \Theta \left[\rho_c-\rho_B(R)\right].
\end{eqnarray}
From Fig.\ref{Fig:3}, it is found that the behavior of bound-state wave functions $\Phi(R)$ of $^{102}$Sn and $^{102}$Te is quite different, and sensitive to their contributing single-particle states. On the contrary, the behavior of $\Phi(R)$ of $^{210}$Pb and $^{210}$Po is rather similar. The scattering state wave functions $\chi(R)$ which merge with the continuum of bound state are obtained as combinations of regular and irregular Coulomb functions, see Fig.\ref{Fig:4}. A strong oscillating feature of $\chi(R)$ is exhibited, as a natural result of two-body Coulomb repulsion.

The decay width given as the product of the pre-exponential factor $\nu$ and the exponential factor ${\cal T}$ is calculated by using the values of $\Phi(R)$ and $\chi(R)$ at the separation radius \cite{Xu2016,Xu2017}:
\begin{eqnarray}
\label{width}
\Gamma=\nu\times{\cal T}=\frac{4\hbar^2 \alpha^2}{\mu k}\left | \Phi(R_{\rm sep}) \chi(R_{\rm sep})\right |^2,
\end{eqnarray}
where $\mu=A_{\alpha}A_d/(A_{\alpha}+A_d)$, $\alpha=\sqrt{2\mu(V(R_{\rm sep})-E_{\rm tunnel})}/\hbar$, $k=\sqrt{2\mu Q_{\alpha}}/\hbar$. $A_d$ is the mass number of the core nucleus and $A_{\alpha}$=4. The tunneling energy is $E_{\rm tunnel}=Q_{\alpha}-28.3$ MeV where $Q_{\alpha}$ is the experimental $\alpha$ decay energy in Refs.\cite{AME2016,Te2018}. Then the decay half-life is
\begin{eqnarray}
\label{thalf}
T_{1/2}=\frac{\hbar \ln2}{P_{\alpha} \Gamma}.
\end{eqnarray}

\begin{table*}[htb]
\caption{The $\alpha$-cluster preformation probabilities and half-lives by the quartetting wave function approach.}\label{Tab:2}
\begin{tabular}{|c|c|c|c|c|c|c|}
\hline
Parent &$Z$ &$N$ &$Q_\alpha$ (MeV) &$P_\alpha$     &$T_{1/2}^{\rm calc.}$ (s)  &$T_{1/2}^{\rm expt.}$ (s)\\
\hline
$^{102}$Sn &50 &52 &$/$    &0.0551 &$/$    &$/$    \\
$^{102}$Te &52 &50 &$/$    &0.3718 &$/$    &$/$    \\
$^{104}$Te &52 &52 &4.900  &0.7235 &1.479$\times10^{-8}$ &$<$1.8$\times10^{-8}$\\
\hline
$^{210}$Pb &82 &128 &3.792 &0.0176 &1.777$\times10^{16}$ &3.701$\times10^{16}$\\
$^{210}$Po &84 &126 &5.408 &0.0137 &1.060$\times10^{7}$ &1.196$\times10^{7}$\\
$^{212}$Po &84 &128 &8.954 &0.1045 &3.395$\times10^{-7}$ &2.997$\times10^{-7}$\\
\hline
\end{tabular}
\end{table*}

The computed $\alpha$-cluster formation probabilities and half-lives for $^{104}$Te, $^{212}$Po and their neighbors $^{102}$Sn, $^{102}$Te, $^{210}$Pb and $^{210}$Po are listed in Table~\ref{Tab:2}. At present, there are no experimental decay energy and half-life available for $^{102}$Sn and $^{102}$Te, only their $\alpha$-cluster formation probabilities are predicted in Table~\ref{Tab:2}. For $^{104}$Te, the calculated $\alpha$-decay half-life is within the range of experimental upper limit ($<$1.8$\times10^{-8}$ s) \cite{Te2018}. An enhanced $\alpha$-cluster formation probability is found for $^{104}$Te, which agrees very nicely with our empirical result $P_\alpha$=0.73 in Ref.\cite{Xu2006}. The computed $\alpha$-cluster formation probability in $^{102}$Te is also large because the single-nucleon wavefunctions of two contributing protons in $^{102}$Te extend much farther to the surface region compared with $^{102}$Sn. The magnitude of $\alpha$-cluster formation probability in $^{212}$Po is several times larger than those in its neighbors $^{210}$Pb and $^{210}$Po. There exists a sudden change of $\alpha$-decay half-lives from T($^{210}$Po)=1.196$\times10^{7}$ s to T($^{212}$Po)=2.997$\times10^{-7}$ s. This sudden change is a result of shell structure effect across the N=126 major shell, which is very difficult to reproduce in $\alpha$-decay models \cite{Brown,Hatsukawa}. However, it is found from Table~\ref{Tab:2} that the experimental $\alpha$-decay half-lives of $^{210}$Pb, $^{210}$Po and $^{212}$Po are well reproduced by QWFA. This can be considered as a quite important success of our theory.

\section{Summary}

By using the quartetting wave function approach, we present a microscopic calculation of $\alpha$-cluster formation and decay in $^{104}$Te, $^{212}$Po and their neighbors $^{102}$Sn, $^{102}$Te, $^{210}$Pb and $^{210}$Po. An improved treatment of shell structure for the core nucleus is added instead of the rigid Thomas-Fermi rule. It is found that the effective c.o.m. potential for the quartet is quite sensitive to the contributing single-particle wavefunctions. A pocket is still formed for the effective c.o.m. potential after introducing shell model states for the core nucleus. Striking shell effects on the $\alpha$-cluster formation probabilities are shown for magic numbers 50, 82 and 126 by using the same NN interaction. An enhanced $\alpha$-cluster formation probability is shown for both $^{104}$Te and $^{212}$Po as compared with their neighbors. The observed data of $\alpha$-decay half-lives are reproduced quite nicely by present QWFA calculations. Several improvements could be made in future. For instance, the gradient terms of the equations for intrinsic motion which appear in the inhomogeneous matter can be included and the spatial extension of the $\alpha$-particle may be considered to improve the local density approximation for the Pauli blocking term. We expect that a better account of gradient effects will lead to an effective potential $W( R)$ for the c.o.m. motion of the quartet where sharp edges are avoided.
\\
\\
$\bf{Acknowledgments}$

We would like to thank Dr. Xin Zhang for helpful discussions. The work is supported by the National Natural Science Foundation of China (Grants No.11822503, No.11575082, No. 11535004, No. 11761161001, No. 11975167), by the Fundamental Research Funds for the Central Universities (Nanjing University), and by the National R$\&$D Program of China (Grants No.2018YFA0404403, No.2016YFE0129300).


\begin{thebibliography}{99}

\bibitem{Mang60} H. J. Mang, Phys. Rev {\bf 119}, 1069 (1960).

\bibitem{Delion09} D. S. Delion, Phys. Rev. C {\bf 80}, 024310 (2009).

\bibitem{DLW} D. S. Delion, R. J. Liotta, and R. Wyss, Phys. Rev. C  {\bf 92}, 051301(R) (2015).

\bibitem{Denisov} V. Yu. Denisov and A. A. Khudenko, Phys. Rev. C {\bf 79}, 054614 (2009).

\bibitem{Buck} B. Buck, A. C. Merchant, and S. M. Perez, Phys. Rev. C {\bf 45}, 2247 (1992).

\bibitem{Mohr} P. Mohr, Phys. Rev. C {\bf 73}, 031301(R) (2006).

\bibitem{cxu06} C. Xu and Z. Ren, Phys. Rev. C {\bf 74}, 014304 (2006).

\bibitem{Royer} G. Royer and R. A. Gherghescu, Nucl. Phys. A {\bf 699}, 479 (2002).

\bibitem{THSR} A. Tohsaki, H. Horiuchi, P. Schuck, and G. R\"opke, Phys.
Rev. Lett. {\bf 87}, 192501 (2001).

\bibitem{RGM} J. A. Wheeler, Phys. Rev. {\bf 52}, 1083 (1937).

\bibitem{GCM} D. L. Hill, J. A. Wheeler, Phys. Rev. {\bf 89}, 1120 (1953).

\bibitem{FMD} M. Chernykh, H. Feldmeier, T. Neff, P. von Neumann-Cosel,
and A. Richter, Phys. Rev. Lett. {\bf 98}, 032501
(2007) and refs. therein.

\bibitem{AMD} Y. Kanada-En'yo, Progr. Theor. Phys. {\bf 117}, 655 (2007) and refs. therein.

\bibitem{Po}
G. R\"opke, P. Schuck, Y. Funaki, H. Horiuchi, Zhongzhou Ren, A. Tohsaki, Chang Xu, T. Yamada, and Bo Zhou,
 Phys. Rev. C {\bf 90}, 034304 (2014).

\bibitem{Te2018} K. Auranen D. Seweryniak, M. Albers, A. D. Ayangeakaa, S. Bottoni, M. P. Carpenter, C. J. Chiara, P. Copp, H. M. David, D. T. Doherty, J. Harker, C. R. Hoffman, R. V. F. Janssens, T. L. Khoo, S. A. Kuvin, T. Lauritsen, G. Lotay, A. M. Rogers, J. Sethi, C. Scholey, R. Talwar, W. B. Walters, P. J. Woods, and S. Zhu, Phys. Rev. Lett. {\bf 121}, 182501 (2018).

\bibitem{Te2019} Y. Xiao, S. Go, R. Grzywacz, R. Orlandi, A. N. Andreyev, M. Asai, M. A. Bentley, G. de Angelis, C. J. Gross, P. Hausladen, K. Hirose, S. Hofmann, H. Ikezoe, D. G. Jenkins, B. Kindler, R. L\'eguillon, B. Lommel, H. Makii, C. Mazzocchi, K. Nishio, P. Parkhurst, S. V. Paulauskas, C. M. Petrache, K. P. Rykaczewski, T. K. Sato, J. Smallcombe, A. Toyoshima, K. Tsukada, K. Vaigneur, and R. Wadsworth, Phys. Rev. C {\bf 100}, 034315 (2019).

\bibitem{AME2016} G. Audi, F. G. Kondev, M. Wang, W. J. Huang, S. Naimi, Chin. Phys. C {\bf 41(3)}, 030001 (2017).

\bibitem{review1} R. G. Lovas, R. J. Liotta, A. Insolia, K. Varga, and D. S. Delion, Phys. Rep. {\bf 294}, 265 (1998).

\bibitem{review2} D. S. Delion, R. J. Liotta, R. Wyss, Phys. Rep. {\bf 424}, 113 (2006).

\bibitem{review3} Chong Qi, R. J. Liotta, R. Wyss, Prog. Part. Nucl. Phys. {\bf 105}, 214 (2019).

\bibitem{Xu2016} C. Xu {\it et al.}, Phys. Rev. C {\bf 93}, 011306(R) (2016).

\bibitem{Xu2017} C. Xu, G. R\"opke, P. Schuck, Z. Ren, Y. Funaki, H. Horiuchi, A. Tohsaki, T. Yamada, and Bo Zhou, Phys. Rev. C {\bf 95}, 061306 (2017).

\bibitem{quartet2017} G. R\"opke, P. Schuck, C. Xu {\it et al.}, J. Low Temp. Phys. {\bf 189}, 383 (2017).

\bibitem{quartet2018} G. R\"opke, AIP Conference Proceedings {\bf 2038}, 020008 (2018).

\bibitem{SPM} A. Bohr, B. R. Mottelson, Nuclear Structure, World Scientific Publishing (1998).

\bibitem{Gurvitz1988}
S. A. Gurvitz, Phys. Rev. A {\bf 38}, 034304 (1988); S. A. Gurvitz and G. Kalbermann, Phys. Rev. Lett. {\bf 59}, 262 (1987).

\bibitem{Batchelder} J. C. Batchelder, K. S. Toth, C. R. Bingham, L. T. Brown, L. F. Conticchio, C. N. Davids, D. Seweryniak, J. Wauters, J. L. Wood, and E. F. Zganjar, Phys. Rev. C {\bf 55}, R2142 (1997).

\bibitem{Tarbert2014} C. M. Tarbert {\it et al.}, Phys. Rev. Lett. {\bf 112}, 242502 (2014).

\bibitem{Fricke1995} G. Fricke, C. Bernhardt, K. Heilig, L. A. Schaller, L. Schellenberg, E. B. Shera, and C. W. de Jager, At. Data Nucl. Data Tables {\bf 60}, 177 (1995).

\bibitem{r100} Z. Ren {\it et al.}, J. Phys. G: Nucl. Part. Phys. {\bf 21}, 691 (1995).

\bibitem{Warda} M. Warda, X. Vi\~nas, X. Roca-Maza, and M. Centelles, Phys. Rev. C {\bf 81}, 054309 (2010).

\bibitem{M3YReview} G. R. Satchler and  W. G. Love, Phys. Rep. {\bf 55}, 183 (1979).

\bibitem{Xu2006} C. Xu and Z. Ren, Phys. Rev. C {\bf 74}, 037302 (2006).

\bibitem{Brown} B. A. Brown,  Phys. Rev. C {\bf 46},  811 (1992).

\bibitem{Hatsukawa} Y. Hatsukawa, H. Nakahara, and D. C. Hoffman, Phys. Rev. C {\bf 42},  674 (1990).



\end{thebibliography}
\end{document}